\documentclass[conference]{IEEEtran}
\usepackage{cite}
\usepackage{amsmath,amssymb,amsfonts}
\usepackage{graphicx}
\usepackage{comment}
\usepackage{mathtools}
\usepackage{algpseudocode}
\usepackage{multirow}
\usepackage{algorithm}
\usepackage{textcomp}
\usepackage[table]{xcolor}
\def\BibTeX{{\rm B\kern-.05em{\sc i\kern-.025em b}\kern-.08em
    T\kern-.1667em\lower.7ex\hbox{E}\kern-.125emX}}

\begin{document}
\IEEEoverridecommandlockouts


\title{HashPIM: High-Throughput SHA-3 \\ via Memristive Digital Processing-in-Memory}

\author{\IEEEauthorblockN{Batel Oved, Orian Leitersdorf, Ronny Ronen, and Shahar Kvatinsky}
\IEEEauthorblockA{\textit{Andrew and Erna Viterbi Faculty of Electrical and Computer Engineering, Technion – Israel Institute of Technology, Israel} \\
batelov@campus.technion.ac.il, orianl@campus.technion.ac.il, ronny.ronen@technion.ac.il, shahar@ee.technion.ac.il}}

\maketitle

\IEEEpubid{\begin{minipage}{\textwidth}\ \\[12pt] \centering \copyright 2022 IEEE. Personal use of this material is permitted. Permission from IEEE must be obtained for all other uses, in any current or future media, including reprinting/republishing this material for advertising or promotional purposes, creating new collective works, for resale or redistribution to servers or lists, or reuse of any copyrighted component of this work in other works.\end{minipage}}


\begin{abstract}
Recent research has sought to accelerate cryptographic hash functions as they are at the core of modern cryptography. Traditional designs, however, suffer from the von Neumann bottleneck that originates from the separation of processing and memory units. An emerging solution to overcome this bottleneck is processing-in-memory (PIM): performing logic within the same devices responsible for memory to eliminate data-transfer and simultaneously provide massive computational parallelism. 
In this paper, we seek to vastly accelerate the state-of-the-art SHA-3 cryptographic function using the memristive memory processing unit (mMPU), a general-purpose memristive PIM architecture. To that end, we propose a novel in-memory algorithm for variable rotation, and utilize an efficient mapping of the SHA-3 state vector for memristive crossbar arrays to efficiently exploit PIM parallelism.
We demonstrate a massive energy efficiency of $\mbox{\boldmath $1,422$}$ Gbps/W, improving a state-of-the-art memristive SHA-3 accelerator (SHINE-2) by $\mbox{\boldmath $4.6\times$}$.
\end{abstract}




\section{Introduction}
\label{sec:introduction}
\IEEEpubidadjcol

As we enter the era of data-intensive computing across many Internet of Things (IoT) devices, cryptography is emerging as a crucial field for secure communication. At the core of this field are cryptographic hash functions~\cite{Cryptography} which are fundamental for tasks such as digital signature generation and verification, key derivation, and pseudo-random bit generation~\cite{NIST, Cryptography}. These functions generalize traditional hashing with additional properties aimed at improving security, such as being very infeasible to invert. An emerging state-of-the-art hash function is Secure Hash Algorithm-3 (SHA-3), which exploits techniques such as sponge construction to enhance security~\cite{NIST}.

While hashing is traditionally implemented via software, hardware accelerators are emerging to provide unparalleled performance using ASICs~\cite{EfficientHardware, RFID, ISCAS, ESSCIRC}, FPGAs~\cite{EfficientHardware, FPGAKeccak}, and memristive memories (e.g., ReRAM)~\cite{SHINE, ReVAMP, VG-MTJ}. Hardware accelerators benefit from the inherent flexibility for bit-wise accesses that does not exist in CPUs and GPUs. Yet, processing units, including hardware accelerators, are subject to the \emph{memory wall}; therefore, when hashing large objects stored in memory (or disk), data transfer becomes the bottleneck~\cite{SHINE}. 

An emerging concept to overcome the memory wall is that of Processing-in-Memory (PIM). The fundamental idea of PIM is to shift the computation into the memory, thereby avoiding the data-transfer between the CPU and memory. Recent techniques for PIM involve using the same physical devices for both \emph{binary} storage and basic \emph{digital} logic gates, via technologies such as digital memristive memories~\cite{NishilLogicDesign, RACER}, DRAM~\cite{DRISA}, SRAM~\cite{ComputeCaches}, and FeFET~\cite{FeFET}. Previous works have sought to utilize this emerging field to accelerate a wide range of applications, including the SHA-3 cryptographic hash function~\cite{SHINE, ReVAMP, VG-MTJ}. Unfortunately, previous SHA-3 designs require specific complex \emph{near-array} periphery that leads to costly data conversion which diminishes the benefit of PIM. Conversely, we seek to design an \emph{in-array} \emph{algorithm} that utilizes the emerging general-purpose memristive Memory Processing Unit (mMPU)~\cite{ShaharCNNA} \emph{without any custom circuitry.}

We focus on a digital memristive PIM architecture~\cite{NishilLogicDesign, RACER, ShaharCNNA} as this technology has vast potential for large-scale efficient PIM; regardless, the proposed algorithms can be generalized to additional PIM techniques and technologies. Memristors~\cite{Memristor} are similar to resistors as they are two-terminal devices, yet they possess a highly unique property: an applied voltage can alter their internal resistance. Therefore, memristors can inherently support storage by representing binary information via their resistance. Interestingly, recent works~\cite{IMPLY, MAGIC, FELIX} have shown that memristors also support basic logic functionality through \textit{stateful logic}~\cite{ShaharCNNA, MemristiveLogic}. Thus, memristors inherently enable both memory and digital logic.

\begin{figure*}[t]
\centerline{
\includegraphics[width=\linewidth]{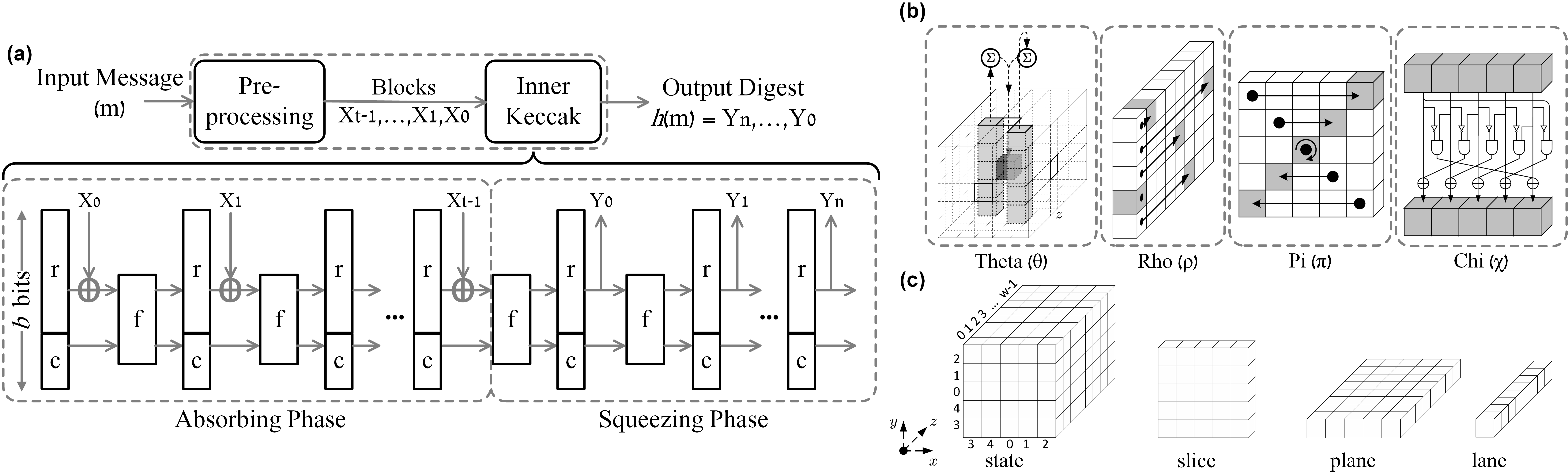}}
\caption{(a) Overview of the SHA-3 algorithm. (b) The Keccak-f round function~\cite{Keccak}. (c) The notation for the state vector and its subvectors~\cite{Keccak}.}
\label{fig:sha3}
\end{figure*}

In this paper, we propose an efficient \emph{in-memory} algorithm for SHA-3, using the mMPU~\cite{ShaharCNNA}, that efficiently exploits the vast potential of PIM. We compare our design to other accelerators and demonstrate superior energy efficiency.

\section{Background}
\label{background}
\IEEEpubidadjcol

\subsection{Secure Hash Algorithm-3 (SHA-3)}
Hash algorithms reduce a large variable-sized input to a small fixed-sized output (hash value) while maintaining a near-uniform output distribution. Cryptography, the field in computer science establishing secure communication, extends the notion of hash functions to \emph{cryptographic hash functions}~\cite{Cryptography} that also possess certain properties which enhance security. For example, these functions must be infeasible to invert and slight modifications to the input drastically change the hash value. The Keccak family of cryptographic hash functions was proposed by Bertoni~\textit{et al.}~\cite{Keccak}, and was later adopted as part of the state-of-the-art SHA-3 standard~\cite{NIST}. This standard proposes four variations for different output lengths; without loss of generality, we discuss SHA3-256 in this paper.

We describe the overall operation of SHA-3, as shown in Fig.~\ref{fig:sha3}(a). The input is a message of size $m$ bits which is padded to be of a length that is a multiple of $r$ and then split into $t$ blocks of size $r$ each (e.g., $r=1088$). The algorithm proceeds with the absorbing phase that initializes an internal state of $b \triangleq r + c$ bits to zero and then iteratively performs an exclusive-or (XOR) with an input block followed by the \verb|Keccak-f| routine. The routine considers the state vector as a $5 \times 5 \times 64$ binary tensor and applies the operations detailed in Algorithm~\ref{alg:keccak} and Fig.~\ref{fig:sha3}(b). The state vector after the final iteration is truncated to receive the overall hash\footnote{Extendable-output-functions continue with additional iterations of Keccak-f as part of the squeezing phase, to generate hash values longer than the state array. The proposed algorithm can also be generalized to such functions.}.

\begin{figure}[t]
\vspace{-20pt}
\begin{algorithm}[H]
\small
\caption{Keccak-f}
\begin{algorithmic}[1]
\Require{State array $A[x][y]$ (for all $x,y \in [0,4]$)}
\Ensure{State array $A[x][y]$ (for all $x,y \in [0,4]$)}
    \For{\texttt{$i_r = 0, ..., 23$}}
        \Statex\hspace{\algorithmicindent}{\textbf{Theta ($\boldsymbol{\theta}$) step:}}
	    \State{$C[x] \gets A[x][0] \oplus \cdots \oplus A[x][4] \hspace{63pt} \forall x \in [0,4]$}
	    \State{$D[x] \gets C[x-1] \oplus (C[x+1] \lll$\textsuperscript{a} $1) \hspace{39pt} \forall x \in [0,4]$}
	    \State{$A[x][y] \gets A[x][y] \oplus D[x] \hspace{75pt} \forall x,y \in [0,4]$}
	    \Statex{\hspace{\algorithmicindent}\textbf{Rho ($\boldsymbol{\rho}$) step:}}
	    \State{$A[x][y] \gets A[x][y] \lll r[x][y]$\textsuperscript{b} $\hspace{56pt} \forall x,y \in [0,4]$}
	    \Statex{\hspace{\algorithmicindent}\textbf{Pi ($\boldsymbol{\pi}$) step:}}
	    \State{$A[y][2x+3y] \gets A[x][y] \hspace{79pt} \forall x,y \in [0,4]$}
	    \Statex\hspace{\algorithmicindent}{\textbf{Chi ($\boldsymbol{\chi}$) step:}}
	    \State{$A[x][y] \gets A[x][y] \oplus (\overline{A[x+1][y]} \land A[x+2][y]) \hspace{18pt} \forall x,y$}
	    \Statex\hspace{\algorithmicindent}{\textbf{Iota ($\boldsymbol{\iota}$) step:}}
	    \State{$A[0][0] \gets A[0][0] \oplus RC[i_r]$\textsuperscript{c}}
    \EndFor
\end{algorithmic}
\textsuperscript{a}The operator $\lll$ denotes a rotation (cyclic shift), \\
\textsuperscript{b}$r[x][y]$ denotes the constants that define the rotation amount~\cite{NIST},\\
\textsuperscript{c}$RC[i_r]$  denotes the constants for each round~\cite{NIST}.\\
$C[x]$, $D[x]$ are intermediates, and indexing is modulo 5.\\
\label{alg:keccak}
\end{algorithm}
\vspace{-20pt}
\end{figure}


\subsection{Memristive Digital Processing-in-Memory (PIM)}

\begin{figure}[t]
\vspace{-15pt}
\centerline{\includegraphics[width=\linewidth]{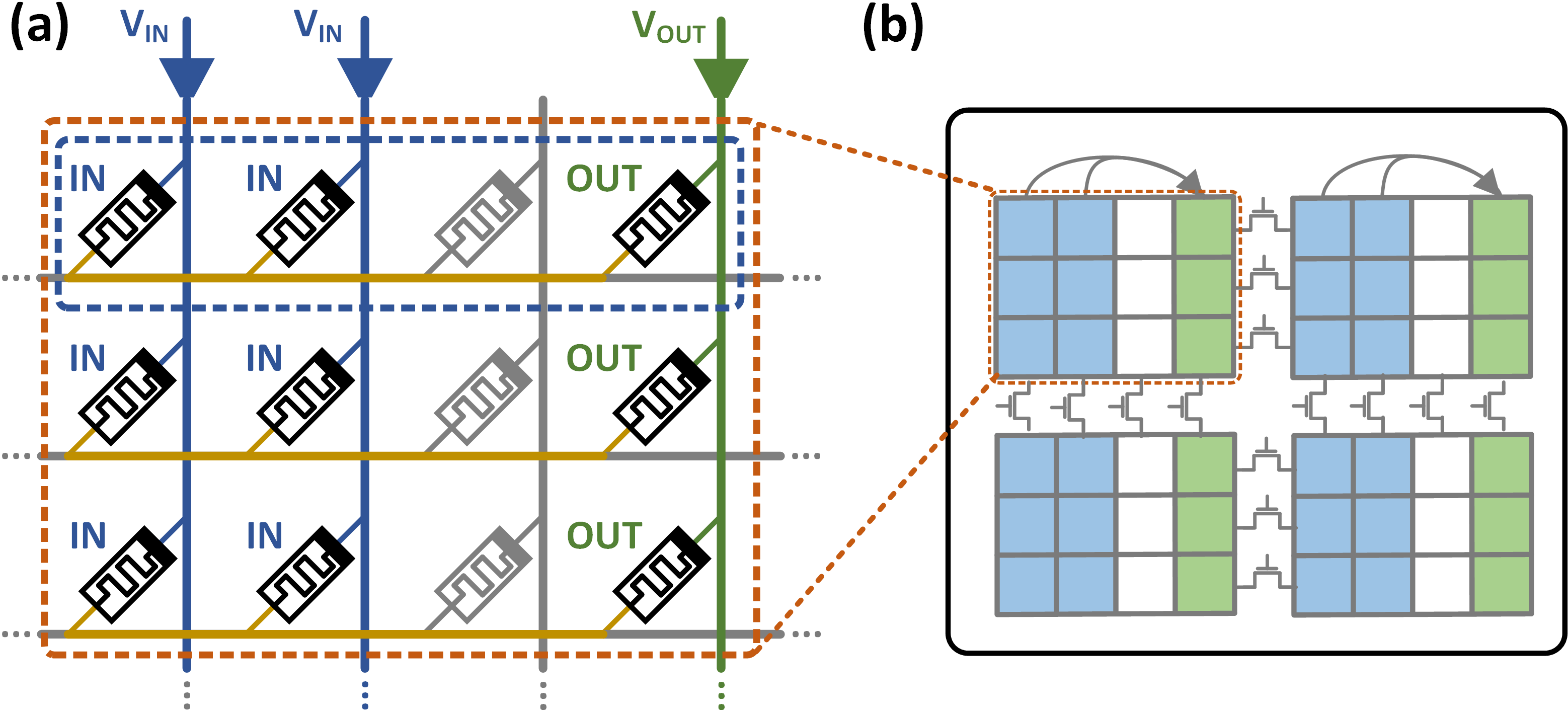}}
\caption{(a) Memristive stateful logic – a single cycle operation across multiple rows simultaneously. (b) Switches creating four independent partitions.}
\label{fig:crossbar}
\vspace{-15pt}
\end{figure}

Memristors~\cite{Memristor} are rapidly emerging as novel physical devices that inherently support both storage and logic functionalities. Memristors are similar to resistors in that they are two-terminal resistive devices, yet they also possess a unique property: a sufficiently-high current can modify their internal resistance. Therefore, memristors can be utilized for binary information storage (e.g., designating high resistance as logical 0 and low resistance as logical 1) as data is written with a relatively high current and read using a relatively low voltage (measuring the current and deriving the resistance). Memristors are typically connected in crossbar array structures, as shown in Fig.~\ref{fig:crossbar}(a). Furthermore, memristors inherently support \emph{stateful logic}~\cite{ShaharCNNA} in the resistive domain which sets the resistance of an output memristor conditional on the states of the input memristors (e.g., performing NOR)~\cite{IMPLY, MAGIC, FELIX}.

The dual functionality of memristors can be exploited towards the memristive Memory Processing Unit (mMPU): a general-purpose memory with massive computational parallelism for bitwise operations, originating from three forms:

\begin{itemize}
    \item \emph{Row/column parallelism:} By applying voltages on bitlines/wordlines of crossbars, we find that several gates can be performed in parallel within the array itself (when their columns/rows are aligned across different rows/columns)~\cite{NishilLogicDesign, RACER}, as shown in Fig.~\ref{fig:crossbar}(a). 
    \item \emph{Partition parallelism:} An overall crossbar may be split into multiple smaller partitions, using transistor switches that divide the bitlines/wordlines, to enable further parallelism~\cite{FELIX, MatPIM, AritPIM}, as shown in Fig.~\ref{fig:crossbar}(b). This enables parallel gates within the same row/column, in addition to the parallelism across multiple rows/columns.
    \item \emph{Crossbar parallelism:} The overall mMPU consists of many crossbar arrays that may operate in parallel~\cite{Bitlet}.
\end{itemize}


\section{HashPIM Architecture}\label{HashPIM}

\begin{figure}[t]
\centerline{\includegraphics[width=\linewidth]{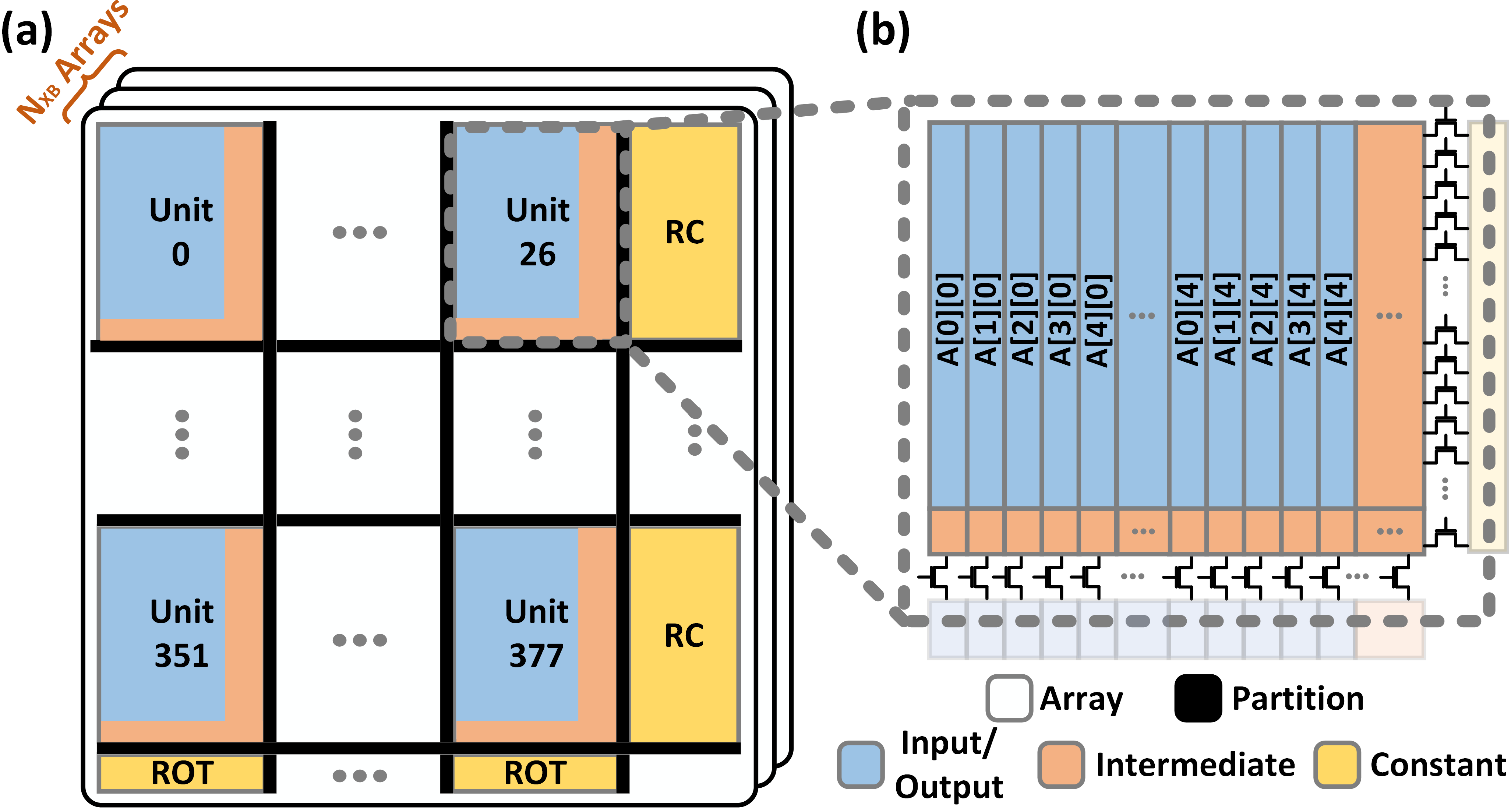}}
\caption{Overview of the proposed architecture for a (a) single SHA-3 crossbar array that holds multiple SHA-3 units. (b) A single SHA-3 unit.}
\label{fig:architecture}
\vspace{-5pt}
\end{figure}

We propose to exploit the vast potential of the mMPU towards an efficient, scalable, and high-throughout SHA-3 algorithm. The benefit of the mMPU over alternative techniques is both due to the ability to efficiently compute precisely where the state vector is stored, and the high efficiency of processing within an array rather than near-array via peripheral circuits.

We detail the proposed architecture within a single crossbar array as the same operations may be performed in parallel across multiple crossbar arrays. Consider a crossbar array as shown in Fig.~\ref{fig:architecture}(a), with size of $1024 \times 1024$ bits, divided into $378$ partitions ($27$ horizontally, $14$ vertically). Each partition (group of $72 \times 37$ memristors) is designated a \emph{SHA-3 unit} and is assigned to compute the SHA-3 hash for a specific message. Thus, the $378$ units enable the parallel computation of SHA-3 on $378$ different messages within the same crossbar array. 

Each SHA-3 unit stores the state-vector corresponding to that message (throughout the computation) and is responsible for applying the \verb|Keccak-f| function on the state-vector. We choose to map the $5 \times 5 \times 64$ state-vector onto $25 \times 64$ memristors (similar to \cite{ReVAMP}) as an analysis of the routines in the \verb|Keccak-f| function revealed that this mapping supports optimal parallelism for the $\theta, \pi, \chi$ and $\iota$ steps. Regarding the $\rho$ step, previous works have been unable to implement this step within the array due to the different rotation amount provided for each \emph{lane} (see Fig.~\ref{fig:sha3}(c)), and have thus instead required to read/write the state-arrays and perform the computation in the periphery. The drawback of such a solution is that the periphery becomes the bottleneck when scaling (e.g., 378 SHA-3 units). Conversely, by extending a recent concept proposed for floating-point operations~\cite{AritPIM}, we demonstrate an efficient in-array design which circumvents such periphery. 

In this section, we detail the proposed design within each SHA-3 unit for the various steps that comprise Algorithm~\ref{alg:keccak}.


\subsection{Theta ($\theta$) Step}
\label{theta}
\begin{figure}[t]
\centerline{
\includegraphics[width=\linewidth, trim={0cm, 0.5cm, 0cm, 0cm}]{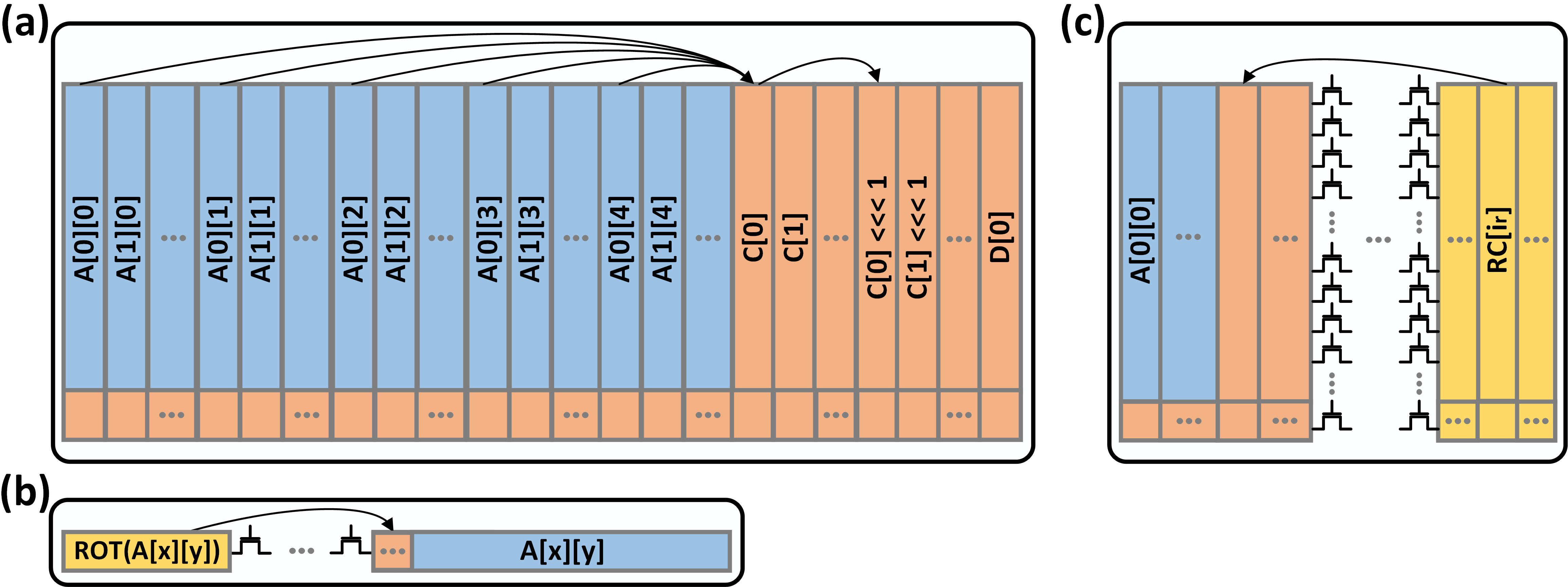}}
\caption{Overview of the operation of the (a) Theta, (b) Rho, and (c) Iota steps. The Pi and Chi steps require only row operations and are thus not drawn.}
\label{fig:steps}
\vspace{-17pt}
\end{figure}

\begin{figure}[t]
\begin{algorithm}[H]
\caption{Variable Rotation}
\begin{algorithmic}[1]
\Require{$N_x$-bit $x_i$, $N_t$-bit $t_i$, for every column $i$.}
\Ensure{$N_x$-bit output $z_i = x_i \lll t_i$ for every column $i$.}
	  \State {$\forall i: z_i \gets x_i$}
      \For{$j = 0, ..., \log_2(N_x)$}
        \State \texttt{$\forall i: z_i \gets mux_{t_i}(z_i \lll 2^j, z_i)$}
      \EndFor
      \State \Return {$z_i$ for every column $i$.}
\end{algorithmic}
\label{alg:rotation}
\end{algorithm}
\vspace{-25pt}
\end{figure}

We start by reducing the state-vector (see Fig.~\ref{fig:sha3}(c)) across the $y$ dimension using exclusive-or (XOR) operations; that is, we compute the XOR of every five columns from the $25 \times 64$ state-vector to result in a $5 \times 64$ vector representing $C[0], \ldots, C[4]$ (see Algorithm~\ref{alg:keccak}) that is stored in the intermediate area (see Fig.~\ref{fig:steps}(a)). We compute $D[0], \ldots, D[4]$ by copying $C[0], \ldots, C[4]$ to $5$ additional intermediate columns (in-row gates), shifting all of the columns once (in-column gates), and then computing the XOR with $C[0], \ldots, C[4]$. Lastly, the original state-vector is updated by computing the XOR of all columns with the relevant lanes from $D[0], \ldots, D[4]$ (e.g., $D[0]$ is XORed with the first five columns in the state-vector).


\subsection{Rho ($\rho$) Step}\label{rho}
At the core of this step is a variable rotation: for each lane in the state vector, we need to rotate that lane by a given number of bits (determined according to $r[x][y]$ see~\cite{NIST}). The difficulty in this \emph{variable} rotation is that, unlike the \emph{constant} shift in $\theta$, every lane may contain a different rotation amount~\cite{DRISA}. This is seemingly an inherent contradiction to the operation of the mMPU as the operation involves data-dependent control flow. 

A similar task of in-array \emph{variable shifting} was recently considered for in-memory floating-point operations~\cite{AritPIM}. Essentially, the variable shift is represented as a sequence of multiplexer operations that are chosen according to a logarithmic-shifter~\cite{ComputerArithmetic} design, thereby converting the control-flow to data-flow. We extend this algorithm to the task of variable rotation (cyclic shifting) in Algorithm~\ref{alg:rotation}. To remain in-place (i.e., not require a copy of the state-vector), we analyze the cyclic dependencies in the rotation and execute them serially with a single slice of redundancy for the storage of the state array.  

We store the constant shift amounts ($r[x][y]$) at the bottom of the crossbar array (ROT in Fig.~\ref{fig:architecture}(a)), shared across all of the units that are aligned vertically (to improve memory utilization for a negligible latency increase). Therefore, as shown in Fig.~\ref{fig:steps}(b), we find that each iteration of Algorithm~\ref{alg:rotation} begins by retrieving the relevant bit for the shift amount from the shared ROT for that vertically-aligned set of units. 


\subsection{Pi ($\pi$) Step}\label{pi}
The step involves reordering the lanes, whereas, at each time, one lane, $A[x][y]$ is copied to the intermediate area along with the target lane $A[y][2x+3y]$ and then the original lane, $A[x][y]$, is copied to the target address. This process is then repeated for $A[y][2x+3y]$ until all of the lanes in the state-array have moved (except for $A[0][0]$ that remains in-place). 


\subsection{Chi ($\chi$) Step}\label{chi}
We convert the expression in Algorithm~\ref{alg:keccak} to a sequence of NOT, NOR, XOR, and COPY operations (using De Morgan's law).
Each \emph{plane} (see Fig.~\ref{fig:sha3}(c)) holds five dependent lanes; hence, all lanes of that plane are inverted and stored at the intermediate area (later to perform the NOR operation with the adequate neighbor lane, see Algorithm~\ref{alg:keccak}). Lastly, each 
lane of that plane is XORed with the NOR result of the neighbor lanes, and is stored back by a COPY operation.


\subsection{Iota($\iota$) Step}\label{iota}
In this step, an XOR operation is performed on lane $A[0][0]$, after creating a local copy of the round constant~\cite{NIST}, $RC[i_r]: i_r \in [0,n_r-1]$ from the RC block (see Fig.~\ref{fig:steps}(c)), to each SHA-3 unit of that row. The new value of $A[0][0]$ is mapped to the original location as the input $A[0][0]$.


\section{Evaluation}\label{evaluation}
The HashPIM architecture and algorithm are evaluated on a $1024 \times 1024$ crossbar array which contains $U_{XB} \triangleq 378$ independent SHA-3 units that operate in parallel (each of size $72 \times 37$). The results are verified by a cycle-accurate simulator~\cite{HashPIMCode} that logically models the crossbar array (with partitions) and provides an interface for in-memory operations, while recording latency, energy and area usage. The results are extended to $N_{XB}$ crossbars.
Overall, one SHA-3 round requires 3,494 cycles ($Latency_{Round}$) and consumes 0.765$nJ$ ($Energy_{Unit}$), assuming $r=1088$ and MAGIC~\cite{MAGIC} gate parameters of 3$ns$ delay ($333MHz$) and $6.4fJ$ energy~\cite{RACER}. The expressions for throughput and power are:
\begin{equation}
Tput_{Unit} = \frac{r}{Latency_{Round}} * f,
\end{equation}
\begin{equation}
Tput_{System} = Tput_{Unit} * U_{XB} * N_{XB},
\end{equation}
\begin{equation}
Power_{System} = \frac{Tput_{System} * Energy_{Unit}}{r}.
\end{equation}

Table~\ref{table:performance} summarizes our results and compares to other SHA-3 accelerators (both CMOS-based and memristor-based). The code repository~\cite{HashPIMCode} includes additional results as well as an explanation for the methodology that led to these results.


\section{Conclusion}\label{conclusion}
This paper demonstrates the vast potential of the mMPU for energy-efficient cryptographic hash algorithms through a case study with the SHA-3 function. We propose a novel variable rotation algorithm and an efficient mapping that exploits the inherent parallelism of the mMPU towards high-throughput SHA-3 execution. This provides a massive throughput per watt of $1,422$ Gbps/W, improving the state-of-the-art by $4.6\times$.

\begin{table}[t]
\caption{Performance Comparison of SHA-3 Hardware Designs}
\vspace{-10pt}
\begin{center}
\begin{tabular}{|c|c|c|c|c|}
\hline
\cellcolor{gray!20}Work&\cellcolor{gray!20}$f$&\cellcolor{gray!20}Tput&
\cellcolor{gray!20}Tput/W&\cellcolor{gray!20}Tput/Area \\
\cellcolor{gray!20}&\cellcolor{gray!20}(MHz)&\cellcolor{gray!20}(Gbps)& \cellcolor{gray!20}(Gbps/W)&\cellcolor{gray!20}(bps/$F^2$) \\
\hline
65nm ASIC~\cite{ISCAS}&1K&48&-&7,619 \\
\hline
SHINE-1~\cite{SHINE}&2K&33.4&263&21,916 \\
\hline
SHINE-2~\cite{SHINE}&2K&54&311&22,227 \\
\hline
\textbf{HashPIM (1 XB)}&\multirow{2}{*}{\textbf{333}}&\textbf{39.2}&\multirow{2}{*}{\textbf{1,422}}&\multirow{2}{*}{\textbf{9,354}} \\
\textbf{HashPIM (2 XB)}&&\textbf{78.4}&& \\
\hline
\end{tabular}
\end{center}
\label{table:performance}
\vspace{-10pt}
\end{table}


\section*{Acknowledgment}
This work was supported by the European Research Council through the European Union's Horizon 2020 Research and Innovation Programe under Grant 757259.

\vspace{30pt}

\bibliographystyle{IEEEtran}
\bibliography{refs}

\end{document}